\documentclass[sigconf,natbib=true]{acmart}
\AtBeginDocument{%
  \providecommand\BibTeX{{%
    \normalfont B\kern-0.5em{\scshape i\kern-0.25em b}\kern-0.8em\TeX}}}

\setcopyright{acmcopyright}
\copyrightyear{2022}
\acmYear{2022}
\acmDOI{XXXXXXX.XXXXXXX}

\usepackage{amssymb}

\newcommand\ie{\textit{i.e.}}
\newcommand\ttt{\texttt}

\begin{document}

\title[An Empirical Study of Uniform-Architecture Knowledge Distillation in Document Ranking]{An Empirical Study of Uniform-Architecture Knowledge Distillation in Document Ranking}
\author{Xubo Qin}
\affiliation{%
 \institution{JD.com}
 \country{China}
}
\email{qratosone@live.com}

\author{Xiyuan Liu}
\affiliation{%
 \institution{University of Colorado, Boulder}
 \country{United States}
}
\email{heathclief.liu.physics@gmail.com}

\author{Xiongfeng Zheng}
\affiliation{%
 \institution{Platform and Content Group, Tencent}
 \country{China}
}
\email{peacezheng@tencent.com}
\author{Jie Liu}
\affiliation{%
 \institution{Platform and Content Group, Tencent}
 \country{China}
}
\email{jesangliu@tencent.com}

\author{Yutao Zhu}
\email{yutao.zhu@umontreal.ca}
\affiliation{%
  \institution{DIRO, Université de Montréal}
  \country{Canada}
}
\begin{abstract}
Although BERT-based ranking models have been commonly used in commercial search engines, they are usually time-consuming for online ranking tasks. Knowledge distillation, which aims at learning a smaller model with comparable performance to a larger model, is a common strategy for reducing the online inference latency. 
In this paper, we investigate the effect of different loss functions for uniform-architecture distillation of BERT-based ranking models. Here ``uniform-architecture'' denotes that both teacher and student models are in cross-encoder architecture, while the student models include small-scaled pre-trained language models. Our experimental results reveal that the optimal distillation configuration for ranking tasks is much different than general natural language processing tasks. Specifically, when the student models are in cross-encoder architecture, a pairwise loss of hard labels is critical for training student models, whereas the distillation objectives of intermediate Transformer layers may hurt performance. These findings emphasize the necessity of carefully designing a distillation strategy (for cross-encoder student models) tailored for document ranking with pairwise training samples.
\end{abstract}



\maketitle

\section{Introduction}
Recent years have witnessed great progress of applying deep learning methods to information retrieval tasks~\cite{neuir}. In particular, on document ranking, pre-trained language models (PLM), such as BERT~\cite{bert}, have achieved state-of-the art performance. 
However, because these pre-trained models often have a large number of parameters, they incur an inevitable computational cost and latency during the inference stage~\cite{timemeasure}. This problem will be even severe when deploying pre-trained models in latency-sensitive online ranking tasks. To tackle this problem, numerous PLM-based knowledge distillation (KD) methods~\cite{kd} have been widely studied. The principle of knowledge distillation can be summarized as follows: first, learn a \textit{teacher} model using the labels in training data, and then learn a \textit{student} model using both the training data and the teacher model. Specifically, when training student models, the ground-truth labels from the data are used as \textit{hard} labels, while the output logits of the teacher model are used as \textit{soft} labels. In comparison to directly optimizing the student model with only hard labels, adding soft labels can assist the student model in learning to simulate the behavior of the teacher model, hence improving performance. It is critical to select an appropriate loss function (also known as \textit{distillation objective}~\cite{tinybert}) for learning soft labels. 

Most of existing approaches~\cite{distilbert,minilm,bertpkd,tinybert,contrast} for language model distillation aim at improving the performance of student model in general natural language understanding (NLU) tasks~\cite{glue}. Specially for document retrieval and ranking tasks, some latest approaches are focusing on cross-architecture distillation approaches~\cite{cross_kd,ernie-search} using cross-encoder~\cite{crossencoder} teachers and ColBERT~\cite{colbert} or dual-encoder~\cite{dualencoder} based students. For those cross-architecture approaches, both teachers and students are using pre-trained models in the same scale (\ie~ ERNIE2.0~\cite{ernie} with 12-layer or 24-layer Transformer~\cite{transformer} Encoders). Since the amount of training samples for document retrieval and ranking tasks is usually very large, it may cost thousands of GPU hours to train the large-scaled teachers and students, which is usually not affordable for some commercial search engines. As a result, it's still necessary to explore the uniform-architecture distillation approaches with small-scaled cross-encoder students. 

In this paper, we conduct an empirical study to investigate the effectiveness of different distillation objectives in document ranking tasks. We focus on analyzing the distillation effect \textbf{given a cross-encoder student model with a small-scaled PLM of 4 Transformer layers}. For uniform-architecture distillation in ranking tasks, one of the latest approaches is Simplified TinyBERT~\cite{simplifiedtinybert} (henceforth denoted as STinyBERT) which proposes some simplifications on TinyBERT~\cite{tinybert} in general NLU tasks. A major problem is that the knowledge distillation objectives in both TinyBERT and STinyBERT are based on pointwise training samples, which fit to classification problems. However, in document ranking, pairwise training is widely used and has demonstrated to be more effective~\cite{ltrliu,burges2010ranknet}. Therefore, our first research problem in this paper is how to combine pairwise ranking with knowledge distillation objectives. We propose a straightforward idea of integrating pairwise ranking into STinyBERT and design a pairwise distillation objective. Then, after analyzing the mechanism of pairwise ranking, we consider that in the original STinyBERT, the distillation objectives based on intermediate Transformer layers are superfluous. Our further empirical study confirms our speculation. Based on the experimental results, we simplify the knowledge distillation objectives in STinyBERT, and finally propose a \textbf{PD-BERT} (PD stands for Pairwise Distillation) strategy for knowledge distillation document ranking.

We conduct experiments on the MS MARCO dataset. The experimental results show that our proposed PD-BERT with fewer distillation objectives outperforms existing methods. Our results clearly indicate that to achieve best performance of ranking models, some novel distillation objectives based on pairwise training samples should be carefully designed, and these objectives should be able to represent the relative relations of the pairwise samples.

\section{Related Works}
\label{BERT-ir}
There are multiple architectures for BERT-based ranking models, including cross-encoder~\cite{crossencoder}, dual encoders~\cite{dualencoder} and ColBERT~\cite{colbert}.  Comparing with other architectures, cross-encoder can achieve the best performance, while its inference latency is the highest comparing with the other ones. This issue is critical for commercial search engines. To address this issue, researchers have proposed various knowledge distillation approaches aiming to train a distilled BERT model to boost up inference without losing much performance, and cross-encoder models are usually used as teacher models. 

Typical distillation approaches~\cite{distilbert,minilm,bertpkd,tinybert,contrast} for PLMs are focusing on general natural language understanding (NLU) tasks~\cite{glue}. For document retrieval and ranking tasks, the corresponding approaches can be divided into \textbf{cross-architecture} and \textbf{uniform-architecture} approaches. Given a cross-encoder teacher model with large-scaled PLM, the cross-architecture approaches such as Margin-MSE~\cite{cross_kd} or ERNIE-Search~\cite{ernie-search} are focusing on distilling a student model in ColBERT or dual encoders architectures with a PLM as large as the teacher (such as 12 Transformer layers). While for uniform-architecture approaches, the student model is a cross-encoder with a PLM in smaller scale. Table~\ref{architecture} shows the differences between cross and uniform architecture.

For ranking tasks, existing state-of-the-art approach of uniform-architecture distillation is Simplified TinyBERT (STinyBERT)~\cite{simplifiedtinybert} which is an extension of TinyBERT~\cite{tinybert} for general NLU tasks. Inheriting the spirit of TinyBERT, STinyBERT employs three objectives to learn the weights of some intermediate Transformer layers, including the weights of embedding layers, hidden states, and attention matrices. And comparing with original TinyBERT, STinyBERT merges the two steps for task-specific distillation into one step, and adds a pointwise loss based on ground-truth labels. However, as the training data of ranking task are usually organized as pairwise training samples~\cite{ltrliu}, the effect of existing distillation objectives with pairwise training samples is still unclear.

Different from existing studies, we propose our PD-BERT strategy to adapt the pointwise-based distillation approaches for pairwise samples. We further conduct an empirical study, based on which we propose a simplification on the distillation objectives and obtain better results.
\begin{table}
\normalsize  
\caption{Comparison between Cross \& Uniform Architecture approaches. ``PLM Scale'' denotes the scale of student models' PLMs. ``Arc.'' and ``Enc.'' are the short of ``Architecture'' and ``Encoder''.}  
\label{architecture}
\small
\centering
\begin{tabular}{lll} 
\toprule
Type & Student Arc. & PLM Scale\\ 
\midrule
Uniform-Arc.  & Cross-Enc. (Same as teacher) & Smaller than teacher \\
Cross-Arc.  & ColBERT or Dual-Enc. & Same as teacher\\
\bottomrule
\end{tabular}  
\end{table}
\section{Proposed Method}
\subsection{Background}
Before diving into the detail of our method, we first briefly introduce some background knowledge about pointwise/pairwise ranking and the commonly used distillation objectives in {STinyBERT}.
\subsubsection{Pointwise vs. Pairwise Ranking}
\label{sec:point}
Pointwise and pairwise approaches are two commonly used methods in learning-to-rank framework. \textbf{Pointwise methods} consider a single document at a time in the loss function. They usually train a classifier (with a cross-entropy loss) to measure whether a document is relevant to a query. Then, the final ranking is achieved by sorting the documents according to their predicted scores. In pointwise methods, the score for each document is independent of those for other documents. On the contrary, \textbf{pairwise methods} consider a pair of documents in loss functions. Their target is to learn the relative order of two documents under a same query. In practice, the pairwise methods usually perform better than pointwise methods because predicting relative order is closer to the nature of ranking~\cite{burges2010ranknet,ltrliu}. Specifically, for a query $q$, assuming there is a relevant document $d^+$ and an irrelevant one $d^-$, the pairwise ranking loss can be defined as:
\begin{align}
    \mathcal{L}_{\rm pair} = \max(0, 1-f(q,d^+)+f(q,d^-)),\label{eq:pair}
\end{align}
where $f(q,d)$ is the ranking score between $q$ and $d$. Similar to pointwise methods, the final ranking list can be obtained by sorting the documents according to their ranking scores.

\subsubsection{Knowledge Distillation in {STinyBERT}}
Simplified TinyBERT (\ttt{STinyBERT}) is designed for document retrieval, which applies some simplifications over \ttt{TinyBERT}~\cite{tinybert}. It is distilled from BERT with the following five objectives:
\begin{align}
    \mathcal{L}_1 = \mathcal{L}_{\rm attn} + \mathcal{L}_{\rm hidn} + \mathcal{L}_{\rm emb} + \mathcal{L}_{\rm hard} + \mathcal{L}_{\rm logits}.\label{eq:kd}
\end{align}
The first three objectives are designed for learning the attention weights, intermediate hidden layers, and embedding layers from the teacher model, which can be defined as: $\mathcal{L}_{i} = {\rm MSE}(L_{i}^{T},L_{i}^{S}), i\in\{\rm attn,hidn,emb\}$. $L_{i}^{T}$ and $L_{i}^{S}$ are attention/hidden/embedding matrices of the teacher and student model, respectively. By optimizing these objectives, the student model can make their parameter metrics closer to the teacher model, thus leading to comparable performance.
The last two objectives are defined in standard knowledge distillation. $\mathcal{L}_{\rm hard}$ is a cross-entropy loss computed on the student model's output logits $\mathbf{z}^{S}$ and the label, while $\mathcal{L}_{\rm logits}$ is a soft cross-entropy loss measuring the discrepancy between $\mathbf{z}^{S}$ and the logits of teacher $\mathbf{z}^{T}$. To adapt for distillation, $\mathbf{z}\in\mathbb{R}^2$ as the ranking task is formulated as a binary classification task. The probability of being relevant (\ie, $\mathbf{z}^{S}[0]$) is used as the ranking score $f(q,d)$. More details can be obtained in the original paper of TinyBERT~\cite{tinybert} and STinyBERT~\cite{simplifiedtinybert}.

\subsection{Pairwise Distillation Objective}
\label{loss_func_def}

In original {STinyBERT}, $\mathcal{L}_{\rm hard}$ is defined as a cross-entropy loss, which can be treated as a pointwise method. As introduced in Section~\ref{sec:point}, researchers have demonstrated that pairwise approaches can perform better than pointwise approaches in practice. Therefore, our first research question is: \textit{how can we combine pairwise ranking with knowledge distillation objectives?}

A straightforward idea is replacing $\mathcal{L}_{\rm hard}$ in Equation~\ref{eq:kd} with $\mathcal{L}_{\rm pair}$ defined in Equation~\ref{eq:pair}. However, different from pointwise ranking, pairwise ranking involves two documents (\ie, $d^+$ and $d^-$). To consider such a document pair in knowledge distillation, we propose a pairwise distillation objective as:
\begin{align}
    \mathcal{L}_2 &= \mathcal{L}_{\rm pair} + \mathcal{L}^+ + \mathcal{L}^-, \label{eq:pstiny}\\
    \mathcal{L}^+ &= \mathcal{L}_{\rm attn}^+ + \mathcal{L}_{\rm hidn}^+ + \mathcal{L}_{\rm emb}^+ + \mathcal{L}_{\rm logits}^+, \\
    \mathcal{L}^- &= \mathcal{L}_{\rm attn}^- + \mathcal{L}_{\rm hidn}^- + \mathcal{L}_{\rm emb}^- + \mathcal{L}_{\rm logits}^-,
\end{align}
where $\mathcal{L}^+$ and $\mathcal{L}^-$ are computed on the positive pair ($q,d^+$) and the negative pair ($q,d^-$), respectively. STinyBERT distilled with $\mathcal{L}_{2}$ is denoted as ``STinyBERT+Pairwise''.

In Equation~\ref{eq:pstiny}, we can see there are nine objectives should be computed in total, which is very complex. So, our next research question is: \textit{are all these objectives necessary?}

Original STinyBERT is designed for pointwise ranking. The distillation objectives for different layers or attentions can help the student model obtain parameters similar to the teacher model, so that the student model can output a similar ranking score. Intuitively, this constraint may not be necessary for our pairwise distillation objective, because our model is trained to learn a relative order between two documents rather than compute a specific ranking score for a single document. To validate our assumption, we conduct an empirical study and find that the distillation objectives for intermediate parameters (\ie, $\mathcal{L}_i^+$ and $\mathcal{L}_i^+, i\in\{\rm attn,emb,hidn\}$) even hurt the model's performance (experimental results are given in Section~\ref{sec:allresults}). As a result, we propose a simplification on our pairwise distillation objective as:
\begin{align}
\label{pdbert-equ}
    \mathcal{L}_3 = \mathcal{L}_{\rm pair} + \mathcal{L}^+_{\rm logits} + \mathcal{L}^-_{\rm logits}.
\end{align}
STinyBERT distilled with our proposed $\mathcal{L}_{3}$ is denoted as ``PD-BERT''.

\section{Experiments}

\subsection{Datasets and Implementation Details}
Following previous work~\cite{simplifiedtinybert}, we conduct experiments on a subset of the MSMARCO~\cite{msmarco} passage ranking dataset, where only the first 3.2M samples are used for training. The queries and passages are truncated into the length of 32 and 120. For evaluation, we use the whole development set provided by MSMARCO official. It includes about 7k queries and each query contains 1k passages for re-ranking. We use MRR@10 as the evaluation metric, which is suggested by the official evaluation.

We use PyTorch~\cite{pytorch} and HuggingFace's Transformers~\cite{DBLP:journals/corr/abs-1910-03771} to implement the models. For the teacher model, 
we fine-tune a pretrained BERT-base~\cite{bert} model for one epoch with a learning rate of 3e-6 and batch size of 32. This model has 12 Transformer layers and the hidden sizes of all layers are 768. 
For all student models, we use the general-distilled TinyBERT checkpoint  to initialize the model.\footnote{https://huggingface.co/huawei-noah/TinyBERT\_General\_4L\_312D. Notice that TinyBERT provides checkpoints for both general and task-specific distillation, we use the general one to initialize the student model.} This model is denoted as TinyBERT-L4, which consists of four Transformer layers with the hidden size of 312. 
The temperature for $\mathcal{L}_{\rm logits}$ is fixed as 1. 
Other hyper-parameters and distillation settings are the same as those in original STinyBERT~\cite{simplifiedtinybert}. 
\subsection{Baseline}
We compare our proposed PD-BERT with three groups of methods:

(1) \textbf{Fine-tune methods.} We fine-tune BERT-base and TinyBERT-L4 on the dataset and report their performance. This BERT-base model is used as the teacher model in the following distillation approaches. 

(2) \textbf{Distillation methods.} To validate the effectiveness of our proposed pairwise distillation objectives, we report the performance of original STinyBERT and STinyBERT+Pairwise. Both of these two approaches includes the transformer-based distillation objectives $\mathcal{L}_{\rm attn}, \mathcal{L}_{\rm hidn}, \mathcal{L}_{\rm emb}$. We further report the performance of PD-BERT, which all those transformer-based objectives are ignored as Equation~\ref{pdbert-equ} describes.

(3) \textbf{Variants of STinyBERT.} To investigate the influence of different knowledge distillation objectives, we remove some objectives from Equation~\ref{eq:pstiny} and test the performance of these variants. 


\subsection{Results and Discussion}
\begin{table}
\normalsize  
\caption{Performance of different approaches. }  
\label{overall}
\small
\centering
\begin{tabular}{llc} 
\toprule
Model & Type & MRR$@$10\\ 
\midrule
BERT-base (Teacher)  & Fine-tuned& \textbf{0.359} \\
TinyBERT-L4  & Fine-tuned& 0.324 \\
\midrule
STinyBERT & Distilled & 0.330 \\
STinyBERT+Pairwise & Distilled & 0.334 \\
Margin-MSE & Distilled & 0.335 \\
PD-BERT & Distilled & \textbf{0.340} \\
\bottomrule
\end{tabular}  
\end{table}
\subsubsection{Overall Results}\label{sec:allresults}
Table~\ref{overall} shows the experimental results of different fine-tune methods and knowledge distillation methods. We can see that our proposed PD-BERT outperforms all other baselines, which clearly demonstrates its superiority. Furthermore, we have the following observations: 

(1) All the distillation approaches can outperform TinyBERT-L4 with fine-tuning, minimizing the gap between the student model and the teacher model (BERT-base). Besides, our PD-BERT has achieved the best performance among all those approaches.

(2) Compared with STinyBERT, STinyBERT+Pairwise achieves 0.004 absolute improvement in terms of MRR@10. The difference of these two models is the ranking loss. So, the improvement validates the effectiveness of our proposed pairwise distillation objective. This result is consistent with the findings of previous work that pairwise ranking usually performs better than pointwise ranking~\cite{ltrliu}. 

(3) Intriguingly, compared with STinyBERT+Pairwise, our PD-BERT is trained with less distillation objectives but achieves better performance, indicating that the Transformer-based distillation objectives do harm to the performance when pairwise training samples are used. This confirms our assumption that our pairwise distillation method can loose the constraint on learning intermediate parameters. With more flexible parameter tuning, the student model can adapt better to downstream tasks when pairwise training samples are given.

\subsubsection{Ablation Studies}
\begin{table}
\normalsize  
\caption{Results of ablation study. $\mathcal{L}_{\rm pair}$ is used in all variants, thus being omitted. All distillation objectives are computed on both positive and negative samples, so the mark {$+/-$} is also omitted.}  
\label{ablationobjectives}
\centering
\small
\begin{tabular}{llc}  
\toprule  
Model & Distillation Objectives & MRR$@10$\\ \midrule
STinyBERT+Pairwise & $\mathcal{L}_{\rm logits}, \mathcal{L}_{\rm emb}, \mathcal{L}_{\rm hidn}, \mathcal{L}_{\rm attn}$ & 0.334 \\
$\hookrightarrow$ \textit{w/o} Intermediate & $\mathcal{L}_{\rm logits}, \mathcal{L}_{\rm emb}$ & 0.339 \\
$\hookrightarrow$ \textit{w/o} Embedding & $\mathcal{L}_{\rm logits}, \mathcal{L}_{\rm attn}, \mathcal{L}_{\rm hidn}$& 0.334 \\
$\hookrightarrow$ \textit{w/o} Logits & $\mathcal{L}_{\rm attn}, \mathcal{L}_{\rm hidn},\mathcal{L}_{\rm emb} $& 0.319 \\
PD-BERT & $\mathcal{L}_{\rm logits}$ & \textbf{0.340} \\
\bottomrule
\end{tabular}  
\end{table}
Previous experimental results imply that some knowledge distillation objectives are not suitable for document ranking tasks, especially for pairwise ranking. Therefore, we conduct an ablation study to investigate the influence of different distillation objectives in STinyBERT+Pairwise. The experimental results are shown in Table~\ref{ablationobjectives}. We can see:


First, removing any distillation objectives for intermediate parameters leads to performance improvement. This confirms again our assumption that simulating parameters in intermediate Transformer layers is unnecessary when training with pairwise distillation objective. Second, compared with $\mathcal{L}_{\rm emb}$, adding $\mathcal{L}_{\rm attn}$ and $\mathcal{L}_{\rm hidn}$ has negative effect on model's performance. Since the number of parameters in intermediate layers is much more than that in embedding layer, we speculate that learning too much parameters may disturb the pairwise distillation. Finally, when removing $\mathcal{L}_{\rm logits}$, the model's performance drops significantly. This demonstrates the importance of learning output logits from the teacher model.


\begin{table}
\caption{Performance with different distillation settings. ``Teacher Layer'' indicates which Transformer layer of the teacher model is used for distillation.}  
\small
\centering
\label{ablationmapping}
\begin{tabular}{lcc}  
\toprule  
Model & Teacher Layer & MRR@10\\ \midrule
STinyBERT+Pairwise & {3,6,9,12} & 0.334\\
$\hookrightarrow$ Last 4 layers & {9,10,11,12} & 0.333 \\
$\hookrightarrow$ Last 1 layer & {12} & 0.339 \\
PD-BERT & - & \textbf{0.340} \\
\bottomrule
\end{tabular}  
\end{table}

\subsubsection{Effect of Different Distillation Settings}
Previous studies~\cite{bert-analysis} have analyzed the effect of different layers of BERT in document ranking and reported that BERT relies heavily on the interactions of the last several layers for predicting the relevance between query and document. Similarly, in knowledge distillation, the selection of distillation layers in BERT may also influence the performance. Following previous work~\cite{tinybert}, we test several different distillation settings over BERT layers (they are also known as mapping functions). 
The results are shown in Table~\ref{ablationmapping}, and we have the following findings:

First, distilling with the last four layers cannot bring further improvement. The explanation for this could be that, while the last four layers are considered to be crucial for capturing matching signals between query and document, they relied on the bottom layers as well. Due to the limited number of layers of STinyBERT, they are unable to extract as much information as vanilla BERT. Second, when only distilling the last layer of BERT, STinyBERT+Pairwise perform slightly worse than our PD-BERT. This reflects that the distillation on intermediate layer of BERT is useless. Therefore, our simplification on distillation objectives is straightforward yet effective.


\begin{figure}
    \centering
    \includegraphics[width=\linewidth]{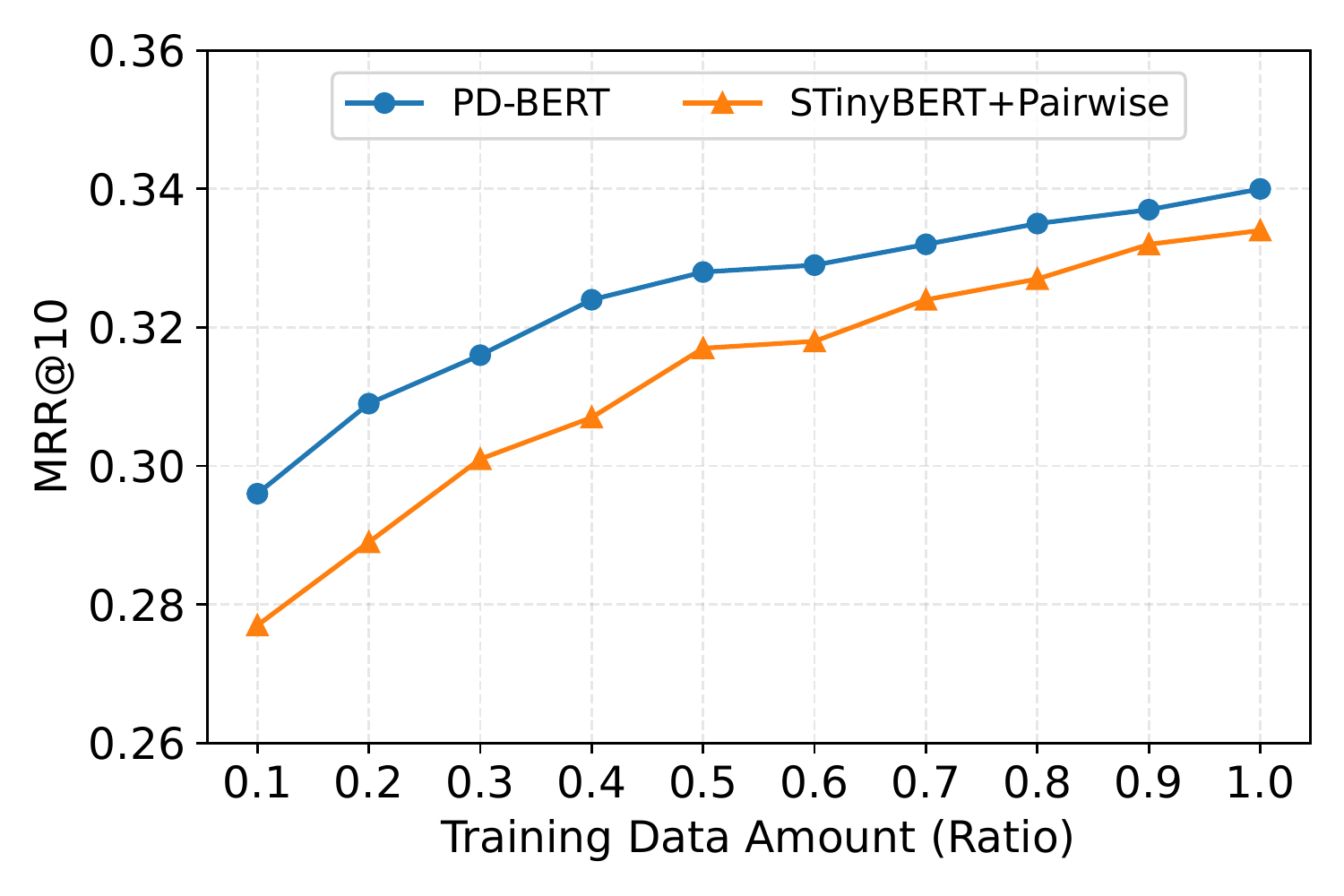}
    \caption{Performance comparison with different amounts of training data.}
    \label{figcompare}
\end{figure}
\subsubsection{Influence of Training Data Amount}
Knowledge distillation is usually influenced by the number of training data, to investigate such impact, 
we test the performance of PD-BERT and PTinyBERT+Pairwise trained with only a proportion of data (from 0.1 to 1.0). The result is shown in Figure~\ref{figcompare}.

It is evident to see that, the performance of both PD-BERT and STinyBERT+Pairwise is increasing with more training data used. This is natural as knowledge distillation benefits from sufficient data. Moreover, we can also observe that the performance difference between STinyBERT+Pairwise and PD-BERT becomes less when providing more training data. A possible explanation is, in pairwise training many candidate documents are repeatedly used in multiple document pairs. In Section~\ref{loss_func_def}, we described that the distillation objectives cannot measure the relative orders between those document pairs. As a result, most of those documents in training set will be redundant for the distillation objectives, which may lead to over-fitting. Compared with the distillation objective on output logits, those aiming at learning intermediate parameters are easier to over-fit and require more unique training documents for tuning. 


\section{Conclusion}
In this paper we investigate the effect of those widely-used objectives of uniform-architecture distillation in ranking tasks. For the student models in cross-encoder architecture, we demonstrate that existing Transformer-based objectives in TinyBERT are not optimal with pairwise training samples. The distillation objectives of intermediate Transformer layers will even do harm to the student model's performance. This may be because the existing distillation objectives are based on pointwise training samples, and the Transformer-based objectives are easier to overfit since there are not enough unique pointwise samples in pairwise training. To fit the ranking task, some kinds of novel distillation objectives should be carefully designed to measure the relative orders in pairwise samples.
\clearpage
\balance
\bibliographystyle{ACM-Reference-Format}
\bibliography{distillation}
\end{document}